\documentclass[sigconf]{acmart}

\usepackage{amsfonts}
\usepackage{algorithmic}
\usepackage{textcomp}
\usepackage{enumitem}
\usepackage{xspace}
\usepackage{pifont}

\newcommand{\niparagraph}[1]{\vspace{2pt}\noindent\textbf{#1}}

\AtBeginDocument{%
  }

\copyrightyear{2025}
\acmYear{2025}
\setcopyright{cc}
\setcctype{by}
\acmConference[ISCA '25]{Proceedings of the 52nd Annual International Symposium on Computer Architecture}{June 21--25, 2025}{Tokyo, Japan}
\acmBooktitle{Proceedings of the 52nd Annual International Symposium on Computer Architecture (ISCA '25), June 21--25, 2025, Tokyo, Japan}
\acmDOI{10.1145/3695053.3731019}
\acmISBN{979-8-4007-1261-6/2025/06}

\begin{document}

\title{{Oaken: Fast and Efficient LLM Serving with Online-Offline Hybrid KV Cache Quantization}}

%
\author{Minsu Kim}
\authornote{Co-first authors who contributed equally to this work.}
\affiliation{
  \institution{KAIST}
  \city{Daejeon}
  \country{Republic of Korea}
}
\email{mskim@casys.kaist.ac.kr}

\author{Seongmin Hong}
\authornotemark[1]
\affiliation{
  \institution{HyperAccel}
  \city{Seoul}
  \country{Republic of Korea}
}
\email{sm.hong@hyperaccel.ai}

\author{RyeoWook Ko}
\affiliation{
  \institution{KAIST}
  \city{Daejeon}
  \country{Republic of Korea}
}
\email{ryeowookko@kaist.ac.kr}

\author{Soongyu Choi}
\affiliation{
  \institution{KAIST}
  \city{Daejeon}
  \country{Republic of Korea}
}
\email{soongyu1291@kaist.ac.kr}

\author{Hunjong Lee}
\affiliation{
  \institution{HyperAccel}
  \city{Seoul}
  \country{Republic of Korea}
}
\email{hj.lee@hyperaccel.ai}

\author{Junsoo Kim}
\affiliation{
  \institution{HyperAccel}
  \city{Seoul}
  \country{Republic of Korea}
}
\email{js.kim@hyperaccel.ai}

\author{Joo-Young Kim}
\affiliation{
  \institution{HyperAccel}
  \city{Seoul}
  \country{Republic of Korea}
}
\email{jy.kim@hyperaccel.ai}

\author{Jongse Park}
\affiliation{
  \institution{KAIST}
  \city{Daejeon}
  \country{Republic of Korea}
}
\email{jspark@casys.kaist.ac.kr}

\renewcommand{\shortauthors}{Kim, Hong, et al.}




\begin{abstract}
Modern Large Language Model (LLM) serving system batches multiple requests to achieve high throughput, while batching attention operations is challenging, rendering \emph{memory bandwidth} a critical bottleneck.
Today, to mitigate this issue, the community relies on high-end GPUs with multiple high-bandwidth memory (HBM) channels.
Unfortunately, HBM's high bandwidth often comes at the expense of limited \emph{memory capacity}, necessitating systems to scale, which reduces core utilization and increases costs.
Moreover, recent advancements enabling longer contexts for LLMs have substantially increased the key-value (KV) cache size, further intensifying the pressures on memory capacity.
To lower the pressure, the literature has explored KV cache quantization techniques, which commonly use low bitwidth (e.g., INT4) for most values, selectively using higher bitwidth (e.g., FP16) for outlier values.
While this approach helps achieve high accuracy and low bitwidth simultaneously, it comes with the limitation that the cost for online outlier detection is excessively high, negating the advantages of quantization.
Inspired by these insights, we propose Oaken, an acceleration solution that achieves high accuracy and high performance simultaneously through co-designing algorithm and hardware. 
To effectively find a sweet spot in the accuracy-performance trade-off space of KV cache quantization, Oaken employs an online-offline hybrid approach, setting outlier thresholds offline, which are then used to determine the quantization scale online.
To translate the proposed algorithmic technique into tangible performance gains, Oaken also comes with custom quantization/dequantization engines and memory management units that can be integrated with any LLM accelerators. 
We built an Oaken accelerator on top of an LLM accelerator, LPU, and conducted a comprehensive evaluation.
Our experiments show that for a batch size of 256, Oaken achieves up to 1.58$\times$ throughput improvement over NVIDIA A100 GPU, incurring a minimal accuracy loss of only 0.54\% on average, compared to state-of-the-art KV cache quantization techniques.

\end{abstract}

\maketitle


%
\section{Introduction}
\label{sec:intro}

\begin{figure}[t]
    \centering
    \includegraphics[width=0.95\linewidth]{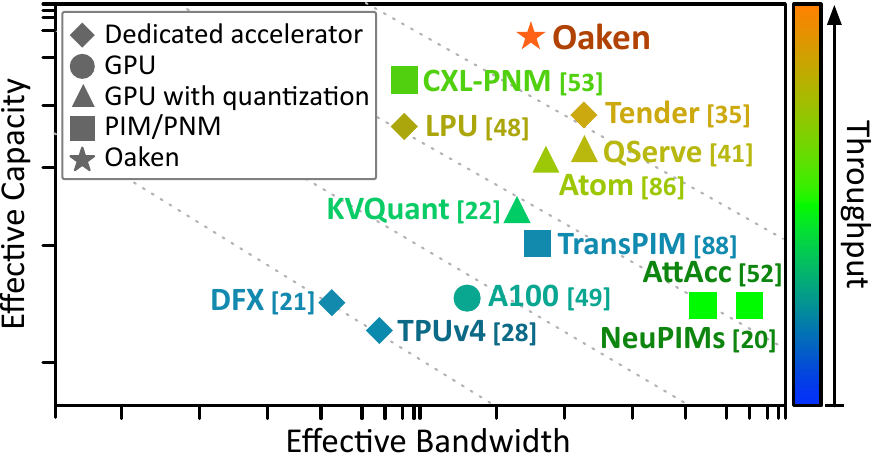}
    \vspace{-0ex}
    \caption{Existing solutions for LLM inference serving systems plotted on the bandwidth-capacity trade-off space. The ``effective bandwidth'' and ``effective capacity'' represent the scale of data that can be transmitted to/from and stored on memory, respectively. We also delineate their corresponding throughput (i.e., tokens/sec) using the colors presented on a 1D heatmap on the right side.}
    \Description{Existing solutions for LLM inference serving systems plotted on the bandwidth-capacity trade-off space. The ``effective bandwidth'' and ``effective capacity'' represent the scale of data that can be transmitted to/from and stored on memory, respectively. We also delineate their corresponding throughput (i.e., tokens/sec) using the colors presented on a 1D heatmap on the right side.}
    \label{fig:tradeoff} 
    \vspace{+2ex}
\end{figure}

The recent advent of large language models (LLMs) has significantly impacted the computing industry.
Almost every sector of the modern economy is exploring the adoption of LLMs, with many already actively using them for various applications.
Most real-world applications rely on hyperscaler-provided LLM serving systems because on-premise LLM deployment requires prohibitive costs.
As LLM inferencing is for multi-tenant environments, the serving systems batch multiple requests to parallelize the inference computation~\cite{yu2022orca,vllm-2023,sheng2023flexgen, cheng2023batch, lina-atc23, dao2022flashattention, nvidia-tensorrt}.
While batching promises a significant throughput boost for operations where operand matrices can be shared across requests on GPUs and other AI accelerators, attention layers in transformer-based LLMs consist of \emph{un}-batchable operations with request-specific, \emph{un}-shareable operands, lowering hardware utilization.
As the \emph{un}-batchable operations cannot exploit on-chip data reuse, they produce enormous memory read, causing \textbf{\emph{memory bandwidth}} to become the key system bottleneck.

Additionally, request batching in LLM inference has another resource implication. 
LLMs generate key-value activation and cache them in memory for computation reuse, often called \emph{KV cache}.
As this KV cache is not shared across different user requests, the large batch size is directly translated into a large KV cache size, requiring massive \textbf{\emph{memory capacity}}.
Furthermore, as KV cache size scales linearly with the sequence length, the recent trend of supporting very long sequences (e.g., 2 million tokens~\cite{longrope}) places even greater pressure on memory capacity.

Consequently, LLM serving systems require both high bandwidth and high capacity to enable fast inferencing.
This resource demand aligns with the immense user demand for LLM services, making it challenging for service providers to build cost-effective LLM serving systems.
Existing solutions often choose to trade-off one resource for the other, as visualized in Figure~\ref{fig:tradeoff}. 
Below, we classify the existing solutions into the following three categories: 
\begin{description}[labelindent=0.0em,nolistsep,leftmargin=1.5em]
\item[(1)] \textbf{Prioritizing \emph{bandwidth} over \emph{capacity}:} 
Currently, using HBM-equipped GPUs is the de-facto standard solution for LLM inference processing~\cite{jouppi2023tpu, medina2020gaudi, nvidia-a100}.
While this approach achieves massive bandwidth, it often compromises capacity, forcing systems to \emph{scale out}, which not only reduces core utilization but also increases the cost of building the system.
\item[(2)] \textbf{Leveraging PIM and/or PNM:}
Even with HBMs, LLM inferencing systems still face bandwidth bottlenecks. 
To address this challenge, recent works have explored the near-data processing (NDP) paradigm, leveraging PIM~\cite{heo2024neupims,attacc-asplos24,ianus-asplos24} and/or PNM~\cite{park2024lpddr}.
While these approaches mitigate the bandwidth bottleneck, their inherent nature requires further reductions in memory capacity, limiting their viability as a fundamental solution.
\item[(3)] \textbf{Exploring LLM quantization strategies:}
One fundamental strategy for jointly addressing the conflicting objectives is to minimize the memory footprint required for LLM inferencing.
To achieve this goal, a large body of prior work~\cite{kim2023squeezellm, hooper2024kvquant, lee2024owq, lin2023awq, kim2023finequant, frantar2022optq, gobo-micro20, chee2023quip, yao2022zeroquant, lin2024qserve, zhao2024atom, liu2024kivi, yuan2023rptq} have recently developed LLM-targeted quantization techniques. 
While these techniques successfully achieve significant reductions in bitwidth, they often prioritize minimizing bitwidth over effectively translating these reductions into practical inferencing speedups.
\end{description}
Alone, none of the solutions is sufficient for building fast and efficient LLM serving systems, which motivates us to develop an acceleration solution, namely Oaken. 
By jointly leveraging algorithmic and hardware techniques, Oaken achieves otherwise unattainable levels of \emph{effective} bandwidth and capacity, resulting in substantially higher throughput than alternatives, as shown in Figure~\ref{fig:tradeoff}.
Oaken comprises (1) \textbf{Algorithmic technique:} an online-offline hybrid KV cache quantization technique, and (2) \textbf{Hardware technique:} the hardware incarnations of the proposed algorithms, including quantization/dequantization engines and memory management units that can be integrated with any existing LLM accelerators.
Oaken makes the following contributions:

\begin{description}[labelindent=0.0em,nolistsep,leftmargin=1.5em]
\item[(1)] \textbf{Online-offline hybrid KV cache quantization: } 
Many of the recently-proposed KV cache quantization techniques commonly use low bitwidth (e.g., INT4) for most values, while selectively using higher bitwidth (e.g., FP16) for outlier values~\cite{hooper2024kvquant, zhao2024atom, lee2024owq, dettmers2022llmint8, guo2023olive, zadeh2022mokey}.
While this approach achieves high accuracy and low bitwidth, the prohibitively high cost of online threshold calculation or mixed-precision computation renders it nearly impractical for real-world use cases. 
Thus, there is a critical need for a cost-effective solution to identify thresholds that distinguish outliers from inliers, enabling the translation of KV cache quantization into tangible speedups.
To achieve this goal, Oaken employs an online-offline hybrid approach, where data-agnostic outlier thresholds are determined at \emph{offline} and subsequently applied to set the quantization scale at \emph{online}.
Furthermore, Oaken introduces a quantization loss mitigation technique that shifts values toward a smaller range, converting outliers into inliers prior to quantization.
Finally, Oaken stores the quantized values by using dense tensors for inliers and fusing sparse outliers into the dense tensors.

\item[(2)] \textbf{Quantization-aware hardware modules for LLM accelerators:} 
We devise custom quantization/dequantization engines and memory management units, which are aware of the proposed quantization algorithm. 
These hardware modules can be integrated with any existing LLM accelerators such as GPUs, NPUs, and LLM-customized ones~\cite{hong2022dfx, park2024lpddr, zeng2024flightllm, hur2023fast}. 
We place these modules in the DMA unit that is commonly present in modern LLM accelerators.
In designing the memory management unit (MMU), the challenge is to achieve the maximal bandwidth, which is close to the physical limit, while effectively laying out the dense and sparse matrices in memory.
We design the MMU with two management tables for dense and sparse data, respectively, to handle virtual-to-physical address mappings and manage the single address space at page granularity. 
This design maximizes memory bandwidth utilization while avoiding fragmentation and burst order issues.
\end{description}

\begin{figure*}[t]
    \centering
    \includegraphics[width=1.0\linewidth]{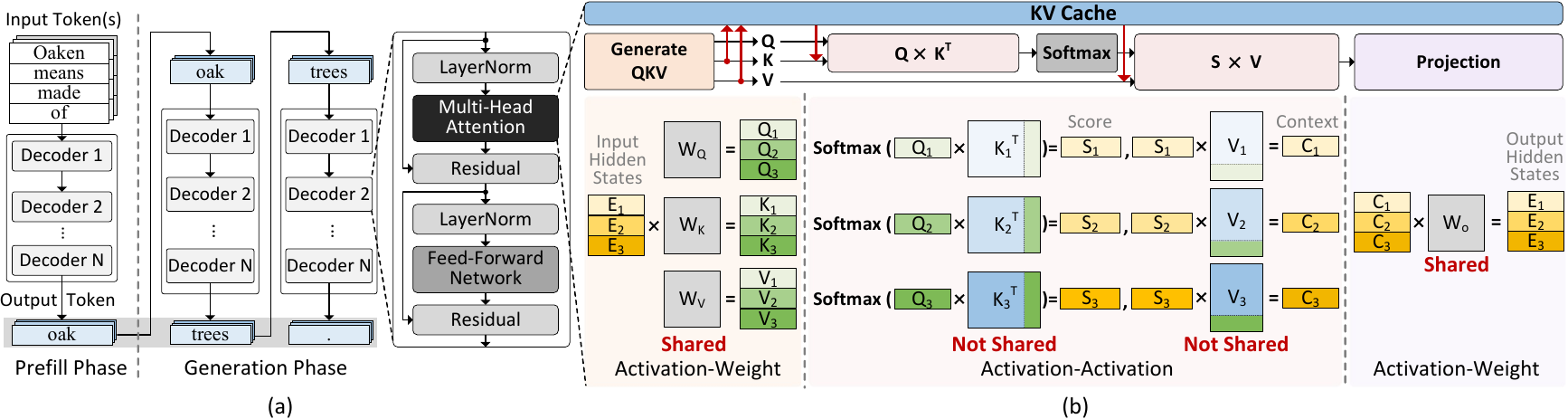}
    \vspace{-3ex}
    \caption{(a) Structure of LLM inference and decoder layer during the prefill and generation phases. (b) Operations in the multi-head attention layer, including activation-weight and activation-activation operations, during the generation phase of batched inference for three requests.}
    \Description{(a) Structure of LLM inference and decoder layer during the prefill and generation phases. (b) Operations in the multi-head attention layer, including activation-weight and activation-activation operations, during the generation phase of batched inference for three requests.}
    \label{fig-back1} 
    \vspace{-0ex}
\end{figure*}

%
To evaluate the effectiveness of Oaken, we use eight different LLMs that include OPT~\cite{zhang2022opt}, Llama2~\cite{touvron2023llama2}, Mistral~\cite{jiang2023mistral}, and Mixtral~\cite{jiang2024mixtralexperts}, with varying sizes.
We use Wikitext, PIQA, Winogrande, and Hellaswag datasets, which are widely used in prior works~\cite{lee2024owq, kim2023finequant, hooper2024kvquant, zhao2024alisa, zhao2024atom, li2024norm, xu2023qalora, cheng2023optimize, lin2024qserve}.
We synthesize the SystemVerilog RTL code of our accelerator in TSMC 28nm technology using Synopsys Design Compiler, which offers the area information of each component in the accelerator.
Our experimental results report that compared to NVIDIA A100 using state-of-the-art KV cache quantization techniques, Oaken offers up to 1.58$\times$ throughput speedup, owing to the bitwidth reduction that reaches up to 70.0\%.  
We achieve this speedup by only introducing a modest 0.87\% accuracy degradation, which demonstrates the algorithmic robustness of our KV cache quantization technique.
Furthermore, Oaken modules incur only 8.21\% area overhead, which is negligible given the significant performance benefits they enable.
These advantages demonstrate that Oaken achieves the dual objectives of bandwidth and capacity, representing an important step toward building fast and cost-effective LLM inference serving systems.
\section{Background}
\label{sec-back}

\subsection{A Primer on LLM Inference}
As illustrated in Figure~\ref{fig-back1}(a), inference of large language model is mainly divided into two phases: the prefill phase and the generation phase~\cite{brown2020language, zhang2022opt, touvron2023llama2, jiang2023mistral, anil2023palm}.
The prefill phase takes input tokens and passes them through decoder layers to generate a single output token.
In the generation phase, the token generated in the previous iteration is used to produce the next token. 
This process is autoregressive, with each iteration generating one token.
LLMs typically consist of multiple decoder layers, with multi-head attention being one of the key operations.
As visualized in Figure \ref{fig-back1}(b), the multi-head attention begins with generating the query, key, and value activations.
The key and value are buffered in the on-chip memory for subsequent operations and are also stored as \emph{KV cache} in the off-chip memory for future iterations.
The query is then multiplied with the transposed key, which is directly fetched from the on-chip memory during the prefill phase, while it is loaded from the off-chip KV cache during the generation phase.
Similarly, subsequent computation using the value requires access to either on-chip memory or the KV cache in the same manner as the key.

\subsection{Batching for High-Throughput LLM Inference}

Batching is a commonly used method to enhance inference throughput in LLM serving systems~\cite{heo2024neupims, vllm-2023, yu2022orca, sheng2023flexgen, agrawal2024taming}. 
It entails processing multiple requests simultaneously and boosts inference throughput by converting memory-bound operations into compute-bound ones through on-chip data reuse.
For instance, in the feed-forward network, weights read from memory can be reused across multiple requests, reducing memory access and increasing throughput.
Figure~\ref{fig-back1}(b) depicts the two types of operations that comprises the multi-head attention layer.
In activation-weight operations, including \textsf{Generate QKV} and \textsf{Projection}, the activations of the requests share the same weights.
This enables on-chip data reuse, reducing memory access and execution time.
However, in activation-activation operations, where the query is multiplied by the transposed key or the score is multiplied by the value, each request requires distinct keys or values, making batched processing challenging.
As a result, batching fails to reduce memory accesses in this case, providing no performance benefits in terms of latency.

\subsection{LLM Quantization}
\label{sec:llm-quantization}
Recently, as demands for LLM inference ever-increasing memory capacity, the community has explored solutions to mitigate this demand through quantization techniques that reduce the bit precision of weights or activations. 
Many of these studies focus on easing the memory pressure caused by model weights, which is particularly effective for accelerating small-batch inference~\cite{kim2023squeezellm, lin2023awq, shao2023omniquant, frantar2022optq, wei2022outlier, yuan2023rptq, wei2023outlier, liu2023qllm}. 
However, these methods achieve limited speedup for larger batches and long sequence lengths due to the KV cache, whose size scales with the batch size and sequence length. 
Approaches to quantize both weights and activations also face similar limitations~\cite{xiao2023smoothquant, lee2023enhancing, sheng2023flexgen, dettmers2022llmint8}. 
To this end, researchers have recently started directing their attention to KV cache quantization.
Some of these works propose quantizing the keys and values of the attention layers on a per-vector basis~\cite{hooper2024kvquant, lin2024qserve, liu2024kivi}, while others propose quantizing the KV cache into multiple vector groups of similar magnitudes, applying channel reordering technique~\cite{zhao2024atom, lee2024tender, lin2024duquant, ashkboos2024quarot}. 
\section{Motivation}
\label{sec:motiv}

This paper focuses on effectively utilizing batched LLM inference to enhance throughput.
To accomplish this, we thoroughly analyze the characteristics of LLM inference, particularly comparing the \emph{un}-batched and batched execution in this section.

\begin{figure}[t]
    \centering
    \includegraphics[width=1.0\linewidth]{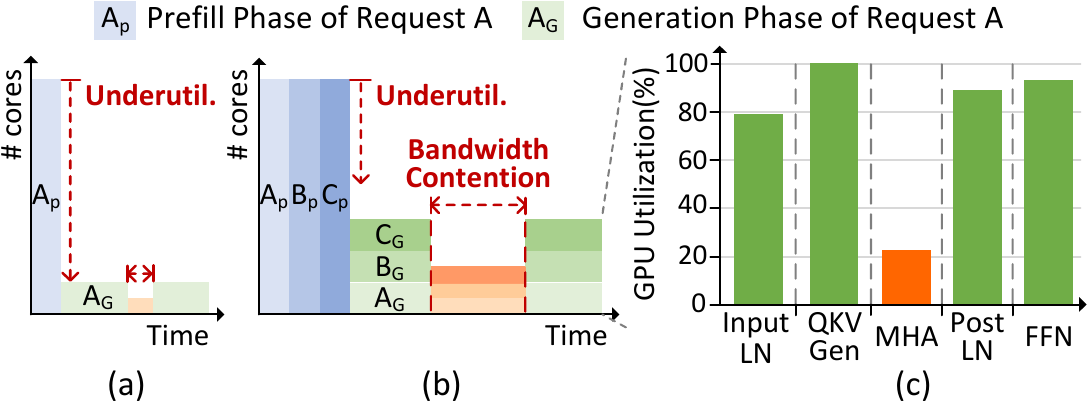}
    \vspace{-3ex}
    \caption{Characteristic analysis of LLM inference for (a) single request and (b) batched multiple requests. (c) Utilization measurement during the generation phase with batched multiple requests using NVIDIA A100 GPU.}
    \Description{Characteristic analysis of LLM inference for (a) single request and (b) batched multiple requests. (c) Utilization measurement during the generation phase with batched multiple requests using NVIDIA A100 GPU.}
    \label{fig-motiv1} 
    \vspace{-0ex}
\end{figure}

\subsection{Characterization of Batched LLM Inference}
\label{sec-motiv-a}
\niparagraph{Single-instance inferencing.}
We analyze the bottlenecks of batched inference by examining accelerator core utilization in both single and batched request scenarios.
Figure~\ref{fig-motiv1}(a) shows the difference in core utilization between the prefill and generation phases when processing a single request.
During the prefill phase, multiple cores are utilized to process multiple input tokens in a parallelized manner.
However, the generation phase processes only one token from the previous iteration, making it highly sequential process that limits parallelization.
Therefore, the accelerator can utilize only a few cores, resulting in high latency in generating all output tokens.

\niparagraph{Batched inferencing.}
Figure~\ref{fig-motiv1}(b) illustrates core utilization for batched inference.
Similar to the single-request scenario, the prefill phase for batched requests is also parallelizable across multiple cores, resulting in high core utilization.
However, the generation phase when processing batched requests still shows underutilization, engendering even longer latency due to bandwidth contention. 
This is because key-value from attention layers cannot be batched and shared across requests, and the accelerator utilizes a limited number of cores per request.
Figure~\ref{fig-motiv1}(c) shows GPU core utilization during the generation phase of Llama2-13B model on an NVIDIA A100 GPU, indicating that underutilization primarily arises from the multi-head attention operations.
\niparagraph{Batching: the double-edged solution.}
A straightforward solution to this issue is to increase the batch size to fully utilize all the cores, but this introduces a new challenge.
The key-value size required for attention operations is proportional to the batch size.
However, as the batch size increases, the memory footprint due to the KV cache also grows, resulting in longer operation latency which is bounded by memory bandwidth.
Recent techniques, such as grouped and multi-query attentions, reduce the memory overhead by shrinking the KV cache size but cannot fully address this issue~\cite{shazeer2019fasttransformer, ainslie2023gqa}.
Our analysis suggests the \textbf{two key observations}: 
\begin{description}[labelindent=0.0em,nolistsep,leftmargin=1.5em]
\item[(1)] While the prefill phase fully utilizes the compute resources, the generation phase often fails to do so.
This imbalance offers an opportunity to enhance throughput with a larger batch size, while it also increases the demand for \textbf{\emph{memory capacity}} due to the KV cache, driving up system construction costs.
\item[(2)] The increased KV cache size due to the aggressive batched processing also causes a high demand for \textbf{\emph{memory bandwidth}} to efficiently compute attention operations, delaying the entire generation phase during batched inference.
\end{description}
%

\begin{figure}[t]
    \centering
    \includegraphics[width=0.95\linewidth]{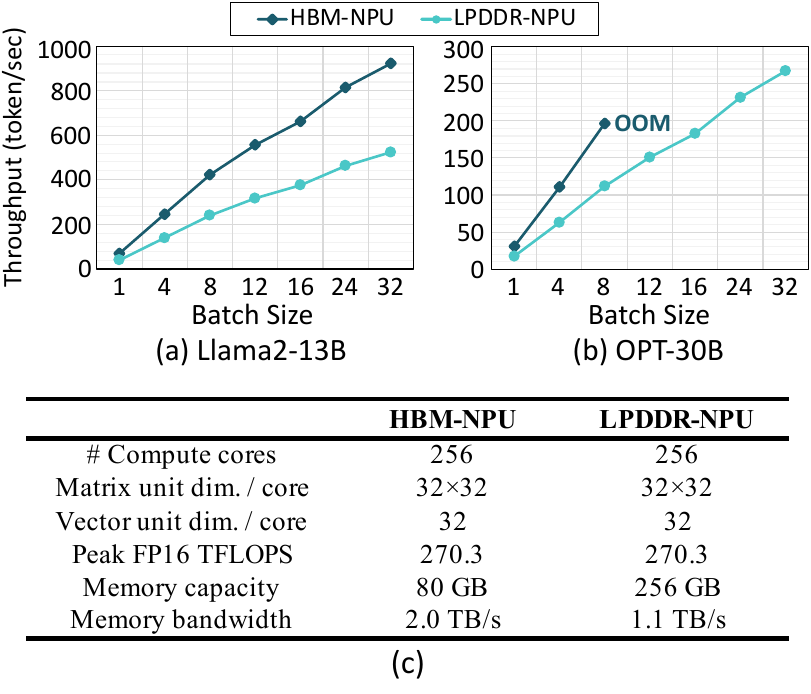}
    \vspace{-1ex}
    \caption{Throughput of accelerators equipped with HBM and LPDDR memory when using (a) Llama2-13B and (b) OPT-30B (OOM refers to ``Out-of-Memory.''). (c) Accelerator specification with HBM and LPDDR memory.}
    \Description{Throughput of accelerators equipped with HBM and LPDDR memory when using (a) Llama2-13B and (b) OPT-30B (OOM refers to ``Out-of-Memory.''). (c) Accelerator specification with HBM and LPDDR memory.}
    \label{fig:memory-motiv-performance} 
    \vspace{-0ex}
\end{figure}

\subsection{Trade-off between Bandwidth and Capacity}
While LLM serving demands both high bandwidth and large capacity, memory technologies exhibit a trade-off between these two resources.
High-bandwidth memory (HBM) sacrifices a substantial portion of capacity to deliver exceptional bandwidth, whereas Low-Power Double Data Rate (LPDDR) DRAM occupies the opposite end of the trade-off space.
Motivated by these observations and insights, we conduct preliminary studies to better understand the performance implications of these conflicting resources on LLM inferencing throughput.
Figure \ref{fig:memory-motiv-performance}(a) and (b) present throughput comparison results for Llama2-13B and OPT-30B models using two accelerator variants with different memory types.
We set both the input and output sequence length to 1K and use an existing LLM-customized accelerator~\cite{moon2024lpu}.
Figure \ref{fig:memory-motiv-performance}(c) lists the specifications of the two accelerators evaluated.
For the smaller LLM model, both accelerators achieve sublinearly-scaling throughput as the batch size increases, with the HBM-based accelerator achieving the highest performance due to its superior bandwidth.
However, for larger LLM models and batch sizes, scaling challenges become evident for the HBM-based accelerator.
In contrast, the LPDDR-based accelerator can accommodate larger batches, demonstrating the best performance.

To summarize, while high memory bandwidth is crucial for achieving high throughput, sufficient memory capacity is also essential to efficiently serve LLMs, particularly with larger models and batches.
Several recent studies also emphasize the importance of memory capacity in LLM inference and suggest introducing large-capacity memory into the LLM accelerators~\cite{park2024lpddr, LLMCompass}.
Despite the industry's growing demand for large-scale LLM inference, current memory technologies struggle to meet both capacity and bandwidth requirements, necessitating a choice.
We explore this trade-off by evaluating our solution, equipped with either LPDDR or HBM memory, across various batched inference scenarios.
%

\begin{figure}[t] 
    \vspace{0in}
    \centering
    \includegraphics[width=1.0\linewidth]{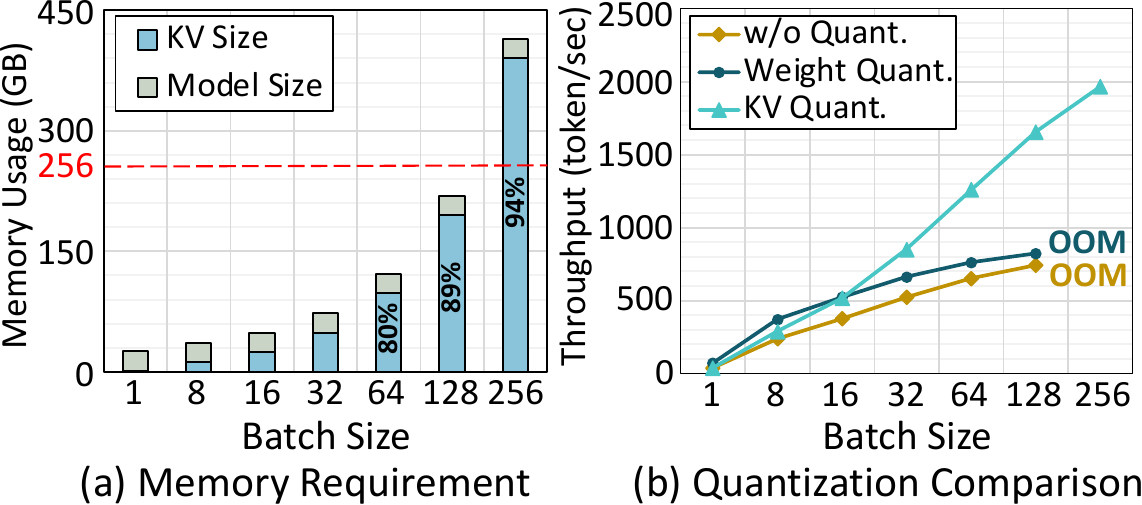}
    \vspace{-2.0ex}
    \caption{(a) Memory usage breakdown to KV cache and model parameters of Llama2-13B model as batch size sweeps from 1 to 256. (b) Throughput comparison among no quantization, weight and KV cache quantization of Llama2-13B model inference. Experiment is conducted with LPDDR-NPU.}
    \Description{(a) Memory usage breakdown to KV cache and model parameters of Llama2-13B model as batch size sweeps from 1 to 256. (b) Throughput comparison among no quantization, weight and KV cache quantization of Llama2-13B model inference. Experiment is conducted with LPDDR-NPU.}
    \label{fig:kv-motiv-performance} 
    \vspace{-0ex}
\end{figure}

\subsection{KV Cache Quantization}
\label{sec:kvcache-background}
Leveraging the memory that offers either sufficient bandwidth or capacity, combined with scaling out the system, can help address the bottlenecks in batched inference.
However, this approach incurs substantial system-building costs and severe resource underutilization, making it neither a fundamental nor a sustainable solution.
\niparagraph{Algorithmic approach to the memory wall.}
To tackle the memory wall challenge, the community has explored algorithmic approaches to reduce the demands on both conflicting resources fundamentally.
Quantization is one such direction, widely recognized for its ability to alleviate both capacity and bandwidth bottlenecks simultaneously.
As discussed in Section~\ref{sec:llm-quantization}, many efforts focus on weight quantization and reducing the computation workload to accelerate LLM inference~\cite{lee2024owq, dettmers2022llmint8, frantar2022optq, park2023lutgemm, yuan2023rptq, xiao2023smoothquant}.
However, our experimental results show that weight-only quantization has a limited impact on addressing the memory wall in the batched scenarios.

\niparagraph{Limitations of weight-only quantization techniques.}
Figure~\ref{fig:kv-motiv-performance}(a) illustrates the memory capacity requirements for model weights and KV cache as batch size increases.
While the memory usage for weights remains constant, the KV cache size grows, eventually dominating the entire device memory.
Figure~\ref{fig:kv-motiv-performance}(b) compares the performance between 4-bit weight quantization and 4-bit KV cache quantization.
The result shows a minimal performance improvement from weight quantization, showing its ineffectiveness in addressing the memory wall for batched LLM inference.
However, KV cache quantization delivers larger performance gains, demonstrating its effectiveness in relieving memory pressure.

\niparagraph{Limitations of existing KV cache quantization solutions.}
Recently, a large body of prior works has explored the KV cache quantization methods~\cite{zhao2024atom,lin2024qserve,lee2024tender,liu2024kivi,ashkboos2024quarot}.
While these works have pioneered a novel research direction, they are constrained by substantial runtime overhead to enable quantization, compromises in accuracy for faster quantization, or a combination of both.
QServe~\cite{lin2024qserve} and Atom~\cite{zhao2024atom} reorder key-value channels by applying the transformation matrix adopted in RPTQ~\cite{yuan2023rptq}, while QServe also handles outliers by applying the scaling matrix introduced by SmoothQuant~\cite{xiao2023smoothquant}.
Tender~\cite{lee2024tender} performs channel reordering via indirect indexing and groups key-value channels with similar magnitude.
These solutions have a limitation in that they come with accuracy degradation due to their low quantization granularity and suffer from additional runtime overhead of channel reordering.
KIVI~\cite{liu2024kivi} and KVQuant~\cite{hooper2024kvquant} propose to use per-vector mixed-precision quantization based on the insight that each KV channel exhibits a distinct pattern in the magnitude of its values.
These methods minimize accuracy loss by isolating outlier channels from the rest, but they incur substantial overhead from sorting operations or mixed-precision computations, which largely offsets the performance gains of quantization.
These limitations in existing solutions emphasize the need for a KV cache quantization technique that achieves low bitwidth without compromising accuracy.
Furthermore, such low bitwidth must translate into tangible throughput improvements at the hardware level.
To meet these demanding requirements, we propose an algorithm-hardware co-designed acceleration solution for efficient batched LLM inference, Oaken.
%


\section{Oaken's KV Quantization Algorithm}
Our quantization algorithm is driven by three primary empirical properties we obtain from observing the value distribution of the KV cache during LLM inference.  
Building upon the three observed properties, we incorporate three algorithmic techniques into Oaken's KV cache quantization.
Below, we will first discuss the empirical analysis results for KV cache distribution and then describe the three techniques one by one. 

\begin{figure*}[t] 
    \centering
    \includegraphics[width=1.0\linewidth]{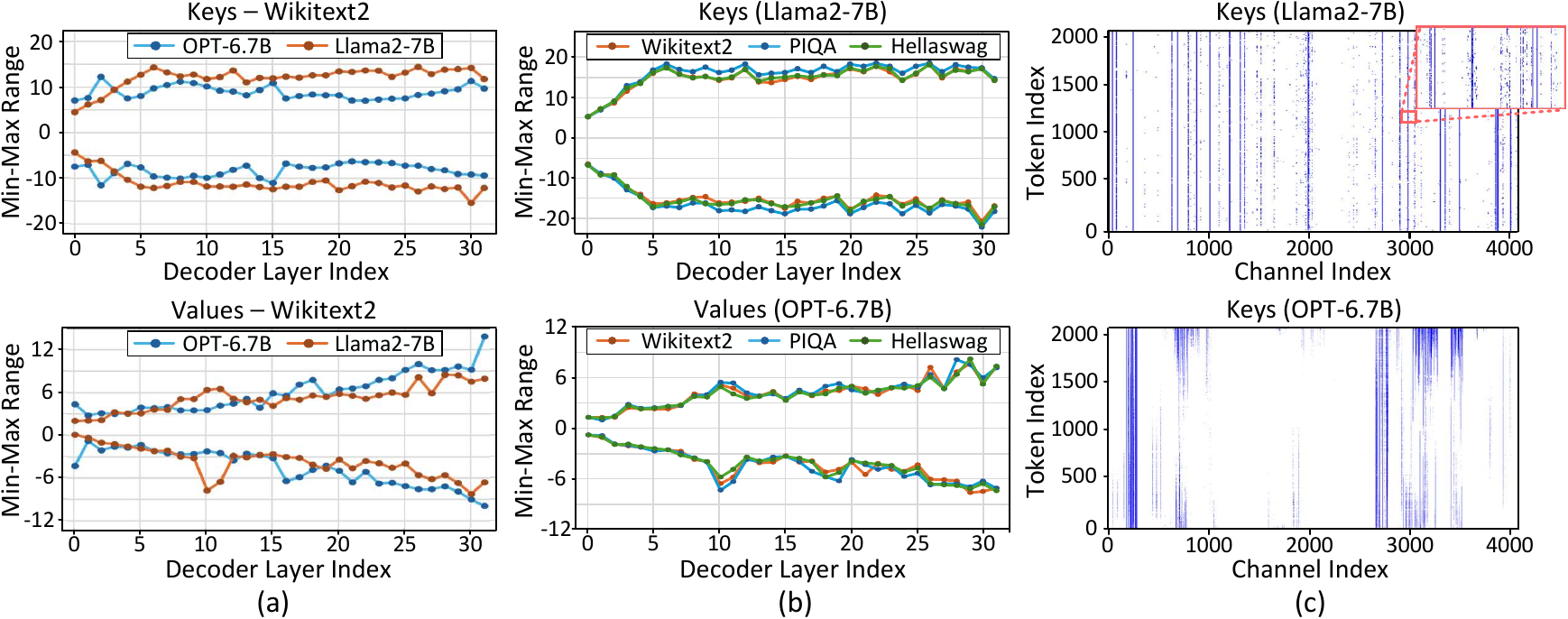}
    \vspace{-3ex}
    \caption{(a) Range of KV values from each decoder layer of OPT-6.7B and Llama2-7B models using Wikitext2 dataset. (b) Range of KV values from each decoder layer of Llama2-7B model using Wikitext2, PIQA, and Hellaswag dataset. (c) Distribution of the top 4\% keys from the 6th decoder layer of Llama2-7B and OPT-6.7B models using Wikitext2 dataset.}
    \Description{(a) Range of KV values from each decoder layer of OPT-6.7B and Llama2-7B models using Wikitext2 dataset. (b) Range of KV values from each decoder layer of Llama2-7B model using Wikitext2, PIQA, and Hellaswag dataset. (c) Distribution of the top 4\% keys from the 6th decoder layer of Llama2-7B and OPT-6.7B models using Wikitext2 dataset.}
    \label{fig:kv-distribution} 
    \vspace{-0ex}
\end{figure*}

\subsection{Observations in KV Distribution}
\label{sec:kv-distribution}
Existing LLM quantization methods often suggest that dealing with large values has a significant impact on model accuracy~\cite{dettmers2022llmint8, zadeh2022mokey, guo2023olive, hooper2024kvquant, lee2024owq, xiao2023smoothquant, zhao2024atom}.
Moreover, other prior works suggest that small values near zero can vanish due to underflow during quantization, leading to larger error~\cite{lee2023enhancing, jin2022fnet, chmiel2024accurateneuraltraining4bit, ahmadian2023intriguing}.
These observations underscore the importance of analyzing the distribution and characteristics of the quantization targets.
We examine the value distribution of the KV cache across several LLMs and datasets and derive key insights for designing effective quantization techniques.
\niparagraph{Observation 1.} 
Figure~\ref{fig:kv-distribution}(a) presents the minimum and maximum range of KV cache values for each decoder layer across various LLMs using the Wikitext2 dataset.
Notably, the magnitude of keys and values varies across models and among decoder layers within each model.
These variations are distinctive properties of each model and decoder layer, driven by differences in their model weights.
From this observation, we gain the insight that the quantization factor should be determined separately for each model and its individual decoder layers.
\niparagraph{Observation 2.}
Figure~\ref{fig:kv-distribution}(b) shows the minimum and maximum range of KV values when using Llama2-7B model with Wikitext2, PIQA, and Hellaswag datasets.
We see that the range of KV cache values remains consistent across these datasets.
This observation implies there is no need for the quantization factor to be tailored to the input sequences but only a global quantization factor per layer.

\niparagraph{Observation 3.}
Prior works have identified a pattern in the magnitude of channel within the KV cache and proposed a per-vector or per-vector-group quantization technique to leverage this pattern~\cite{hooper2024kvquant, lin2024qserve, liu2024kivi, zhao2024atom}.
Figure~\ref{fig:kv-distribution}(c) presents the distribution of the top 4\% values of the key from the sixth layer of the Llama2-7B and OPT-6.7B models.
The distributions exhibit multiple vertical lines, indicating that high-magnitude values are concentrated in specific channels.
This pattern in the channel magnitudes, aligning with previous observations, suggests the need for per-vector or per-vector-group quantization.
However, this result also reveals exceptions to this pattern, appearing as discontinuous lines and dots, which cause an accuracy drop when using only a single quantization scale per vector or vector group.
Based on this observation, we introduce multiple quantization groups within a channel to maintain accuracy, splitting the key-value vector into groups based on the magnitude of each element.
In summary, we derive the following \textbf{three insights} for designing Oaken's quantization technique:

\begin{itemize}[itemsep=0pt, parsep=0pt, topsep=0pt, partopsep=1pt]
    \item Oaken should determine the quantization factor separately for each model and decoder layer.
    \item Oaken can employ a common scaling factor regardless of input prompts, showing its insensitivity to data patterns.
    \item Oaken should use multiple quantization groups segmented by the magnitude of the values within each vector.
\end{itemize}

\begin{figure*}[t]
    \centering
    \includegraphics[width=1.0\linewidth]{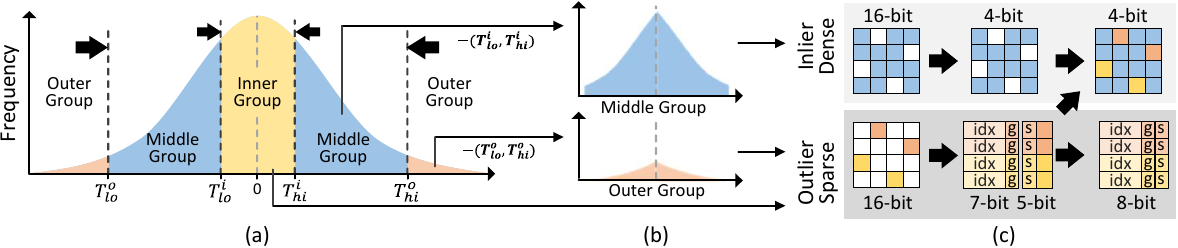}
    \vspace{-4.0ex}
    \caption{Oaken's quantization algorithm consisting of three components: (a) threshold-based online-offline hybrid quantization, (b) group-shift quantization, and (c) fused dense-and-sparse encoding.}
    \Description{Oaken's quantization algorithm consisting of three components: (a) threshold-based online-offline hybrid quantization, (b) group-shift quantization, and (c) fused dense-and-sparse encoding.}
    \label{fig-algorithm}
    \vspace{-0.0ex}
\end{figure*}

\subsection{Algorithm Overview}
We propose a quantization algorithm designed to improve performance by maximizing the compression ratio of the KV cache while minimizing quantization loss, based on the observations and insights from Section~\ref{sec:kv-distribution}.
Figure~\ref{fig-algorithm} illustrates the overall flow of Oaken's quantization algorithm.
Oaken's quantization technique consists of three components:
(1) Threshold-based online-offline hybrid quantization that separates and quantizes inlier and outlier values,
(2) Group-shift quantization that quantizes values with larger magnitude, and
(3) Fused dense-and-sparse encoding that minimizes capacity overhead due to sparse matrices. 
The following sections introduce a detailed design of each component.

\subsection{Threshold-based Online-Offline Hybrid Quantization}
\label{sec:group-quant}
Oaken minimizes quantization loss by isolating \emph{outlier} values that are either exceptionally large or exceptionally small compared to typical \emph{inlier} values.
We propose a threshold-based online-offline hybrid quantization method for more fine-grained grouping.
Oaken separates the \emph{per-token} KV vector into three quantization groups: \emph{outer}, \emph{middle}, and \emph{inner} (Figure~\ref{fig-algorithm}(a)).
The \emph{middle} group consists of inliers where most KV values belong, while the \emph{outer} and \emph{inner} groups consist of outliers, with large and small magnitudes, respectively.
Oaken prevents the quantization scale from being skewed due to large-magnitude outliers by isolating the outer group, and ensures the small-magnitude outliers do not vanish during quantization by isolating the inner group~\cite{lee2023enhancing, jin2022fnet, chmiel2024accurateneuraltraining4bit, ahmadian2023intriguing}.
\niparagraph{Offline outlier threshold profiling.}
To separate outliers from inliers, the \textsf{topK} operation is typically used to maintain a constant ratio of outliers~\cite{hooper2024kvquant}.
While this approach results in minimal quantization loss, the \textsf{topK} operation, essentially a sorting with a time complexity of $O(n\log n)$, introduces significant overhead when performed during inference, degrades end-to-end performance.
Oaken employs offline outlier threshold profiling, leveraging the consistent characteristics in the distribution of KV cache discussed in Section~\ref{sec:kv-distribution} to avoid expensive online operations.
The criteria for splitting the KV cache into three groups are based on four thresholds determined through offline threshold profiling: $T_{lo}^o$, $T_{lo}^i$, $T_{hi}^i$, and $T_{hi}^o$.
Using these thresholds, the outer group, middle group, and inner group are defined as follows:
\begin{equation}
\begin{gathered}
\label{eq:group_definition}
G_o = \{x\;|\;
x < T_{lo}^o
\;\mathrm{or}\;
T_{hi}^o < x\}, \\
G_m = \{x\;|\;
T_{lo}^o \leq x < T_{lo}^i
\;\mathrm{or}\;
T_{hi}^i < x \leq T_{hi}^o\}, \\
G_i = \{x\;|\;
T_{lo}^i \leq x \leq T_{hi}^i\}
\end{gathered}
\end{equation}
Oaken performs approximately a hundred offline inferences with sample input prompts to gather distribution information from the KV cache of each decoder layer.
The four group thresholds are extracted during the profiling process from the KV cache of each inference run using \textsf{topK} operations, and their averages are computed for each decoder layer.
These statistics are then used to establish group thresholds for the KV cache.
As discussed in Section~\ref{sec:kv-distribution}, Oaken's offline profiling should be performed separately for each model, whereas the KV cache distribution and the profiling are independent of both the profiling dataset and future inputs.
It effectively minimizes accuracy loss with only a small number of inferences, ensuring minimal and one-time overhead.
\niparagraph{Uniform quantization.}
Oaken adopts uniform quantization, where the scaling factor $\sigma$ is calculated using only simple statistics to minimize hardware complexity:
\begin{equation}
\label{eq:scaling_factor}
\sigma = \frac{2^m-1}{\mathrm{Max} - \mathrm{Min}}
,
\end{equation}
Where $m$ is the bitwidth of the quantized value, and $\mathrm{Max}$ and $\mathrm{Min}$ represent the maximum and minimum of the values to be quantized. 
The uniform quantization function, which converts a value $x$ into its quantized value, is defined using the scaling factor from Eq.~\ref{eq:scaling_factor}:
\begin{equation}
\label{eq:quantization-function}
Q(x)=\text{round}((x - \mathrm{Min}) \times \sigma).
\end{equation}
Oaken finds the minimum and maximum values for each of the three quantization groups and computes the quantization scaling factor for each group dynamically online.

\niparagraph{Online KV cache quantization.}
%
As previously discussed, Oaken performs per-token quantization on the KV cache, focusing only on the key-value vector newly generated in each attention layer.
Oaken dynamically separates KV cache values into three groups using four group thresholds obtained from offline profiling.
It then retrieves the minimum and maximum values for each quantization group, calculates the scaling factor online, and quantizes the value within each group accordingly.

\subsection{Group-Shift Quantization}
\label{sec:group-shift}
Oaken's threshold-based online grouping effectively mitigates information loss by splitting values into three groups: \emph{outer}, \emph{middle}, and \emph{inner} group.
However, this approach poses a new challenge.
When quantizing the outer group, whose values have large magnitudes, directly applying uniform quantization results in information loss.
Previous works have addressed this by using mixed precision (e.g., \texttt{FP16}) for outliers, distinct from the precision used for inliers (e.g., \texttt{INT4})~\cite{hooper2024kvquant, zhao2024atom, kim2023squeezellm, lee2024owq}.
However, using mixed precision for outliers introduces sparsity, incurring a storage cost of 23 bits per entry: 16 bits for the value, 6 bits for the index, and 1 bit to indicate the group.
Dealing with sparse, mixed-precision outliers also requires additional hardware modules, which adds complexity to the accelerator.
While these overheads are negligible when the outlier fraction is small, they become substantial as the proportion increases.
Moreover, this extra complexity makes it challenging to explore the sweet spot in the accuracy-performance trade-off by adjusting the group thresholds.
One straightforward way to address this issue is to quantize outliers as well.
However, as discussed in Section~\ref{sec:kv-distribution}, quantizing outer groups is challenging due to their wide magnitude range.
To address this, we propose group-shift quantization, a hardware-efficient algorithm that minimizes quantization loss while compressing outliers to a lower bitwidth.
\niparagraph{Group-shift algorithm.}
The core idea of Oaken's group-shift algorithm is to shift the entire group using the thresholds obtained from offline profiling to narrow the range of values, making low-bit quantization possible.
For instance, for the outer group, we subtract $T_{hi}^o$ from values larger than $T_{hi}^o$, and subtract $T_{lo}^o$ from values smaller than $T_{lo}^o$.
While the middle group corresponds to inliers, our group-shift algorithm can also be applied in the same manner.
Figure~\ref{fig-algorithm}(b) shows that applying the aforementioned method shifts the distribution of both the outer and middle groups, concentrating them within a narrower range.
As a result, Oaken can quantize groups spanning wide ranges to low bitwidth using group-shift, minimizing the quantization loss.
Note that the group-shift method does not require additional information beyond those obtained from the offline profiling described in Section~\ref{sec:group-quant}.
In summary, Oaken's quantization function $Q_o(x)$, which converts the value $x$ to the quantized one, is defined as follows:
\begin{equation}
Q_o(x) = \begin{cases}
Q(x - T_{hi}^o) & x \in G_o\;\mathrm{and}\;x > T_{hi}^o\\
Q(x - T_{lo}^o) & x \in G_o\;\mathrm{and}\;x < T_{lo}^o\\
Q(x - T_{hi}^i) & x \in G_m\;\mathrm{and}\;x > T_{hi}^i\\
Q(x - T_{lo}^i) & x \in G_m\;\mathrm{and}\;x < T_{lo}^i\\ 
Q(x) & x \in G_i\\
\end{cases}
,
\end{equation}
where $Q(x)$ is a quantization function defined in Eq.~\ref{eq:quantization-function}.
Oaken quantizes middle group into 4-bit, inner and outer groups into 5-bit.
\subsection{Fused Dense-and-Sparse Encoding}
\label{sec:dense-sparse-encoding}
Oaken employs a dense-and-sparse encoding strategy, as proposed in prior works, to efficiently store dense inliers and sparse outliers~\cite{kim2023squeezellm, hooper2024kvquant}.
In Oaken, the middle group, which consists of inliers and makes up the majority of the KV cache, is stored in a dense matrix.
The outer and inner groups, which consist of outliers, are stored using a sparse matrix format, Coordinate List (COO)~\cite{spatula-micro23, sparch-hpca20, sparsep, spacea-hpca21}, with the corresponding elements in the dense matrix being zeroed.
COO format used in Oaken requires extra 6 bits to indicate the location of each value, along with 1 bit to denote the quantization group, and the bits used to represent the value for each entry.
To further reduce capacity overhead, we propose leveraging the zeroed elements in the dense matrices.
These zeroed elements, corresponding to positions originally occupied by outliers in the KV cache, remain unused after separating the KV cache into dense and sparse matrices.
We introduce a \emph{fused dense-and-sparse encoding} method that repurposes these unused 4 bits to store part of the outliers, as illustrated in Figure~\ref{fig-algorithm}(c).
Specifically, four bits of the quantized 5-bit outliers are embedded in the zeroed elements of the dense matrix, while the remaining 6 index bits, 1 group bit, and 1 sign bit are stored in the sparse COO format.
Since the index bits in the COO format already indicate the location of outliers within the dense matrix, a dedicated flag to denote their presence is unnecessary.
Moreover, with each entry in the sparse matrix fixed at 8 bits and memory-aligned, the memory management unit can efficiently handle both dense and sparse matrices on a page basis, the details of which will be described in Section~\ref{sec:compute-core}.
By combining our fused dense-and-sparse encoding strategy with group-shift quantization, we reduce the bitwidth of each outlier entry from 23 to just 8 bits, increasing the compression ratio of the KV cache, while keeping memory alignment.
 
\section{Oaken's Accelerator Architecture} 
\label{sec:arch}

\subsection{Architecture Overview}
\begin{figure}[t]
    \centering
    \includegraphics[width=1.0\linewidth]{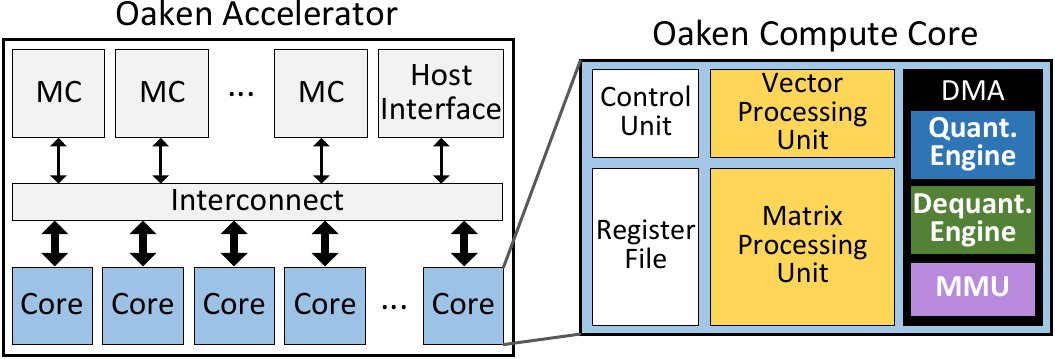}
    \vspace{-2.0ex}
    \caption{Overall Oaken accelerator architecture.}
    \Description{Overall Oaken accelerator architecture.}
    \label{fig-arch} 
    \vspace{-0.0ex}
\end{figure}

Figure~\ref{fig-arch} illustrates the overview of the proposed architecture, which mainly consists of compute cores, memory controllers, a host interface, and an interconnect.
The compute cores are designed to support end-to-end LLM inference operations.
The memory controllers handle device memory to read data, including model parameters, keys, and values, and are also responsible for writing keys and values back to memory.
The host interface employs a PCIe-based connection to communicate with the host system.
This interface also manages the scheduling of incoming requests and distributing them across compute cores for efficient processing.
An interconnect links these components and is optimized to maximize bandwidth utilization during memory read, ensuring efficient data transfer to the compute cores.
This design enables the concurrent use of all memory controllers for reading model parameters and distributing them across the compute cores.
On the other hand, memory writes from the compute cores are less frequent and involve smaller data sizes, reducing bandwidth consumption and simplifying the logic design without compromising performance.

\subsection{Oaken Compute Core} 
\label{sec:compute-core}

\niparagraph{Overall design.}
The main modules of the proposed accelerator are the compute cores, which are adapted from the architecture introduced in LPU~\cite{moon2024lpu} to enable end-to-end LLM inference.
Each compute core consists of a Matrix Processing Unit (MPU) and Vector Processing Unit (VPU) designed to execute LLM inference operations token-by-token.
These processing units are designed to maintain high utilization throughout the entire process while minimizing inefficient logic and ensuring low latency.
MPU is designed to stream weight read from memory to perform efficient matrix-vector multiplication, while VPU handles element-wise operations between matrix-vector computations.
The direct memory access (DMA) unit facilitates data transfer by reading weights from memory to feed the processing units and writing KV cache back to memory.
This DMA unit also incorporates quantization/dequantization engines and a memory management unit (MMU), all of which are critical for implementing the proposed KV quantization technique.
%

\begin{figure}[t]
    \centering
    \includegraphics[width=1.0\linewidth]{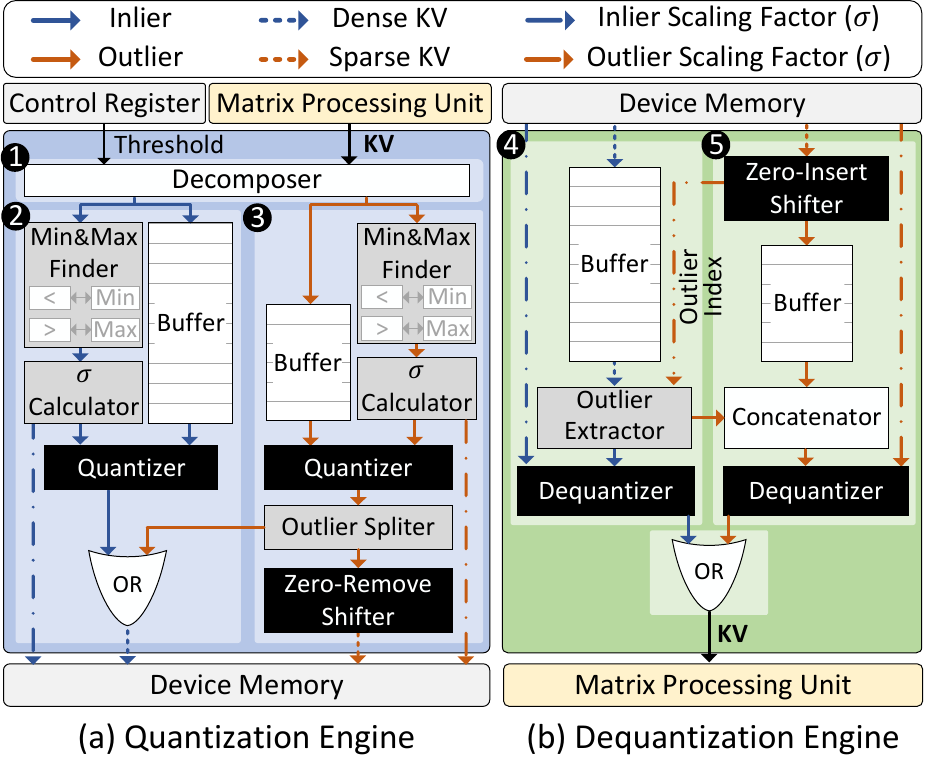}
    \vspace{-3.0ex}
    \caption{Quantization and dequantization engines of Oaken compute core.}
    \Description{Quantization and dequantization engines of Oaken compute core.}
    \label{fig-compression-engine} 
    \vspace{-2.0ex}
\end{figure}

\niparagraph{Quantization engine.}
Figure~\ref{fig-compression-engine}(a) shows the quantization engine in DMA unit, designed to perform online KV cache quantization.
%

%
First, the {\large\ding{172}} \textbf{decomposer module} partitions incoming activations into three quantization groups based on outlier thresholds determined offline.
It then performs group shift for the outer and middle groups by subtracting their streamed thresholds from the values of these groups.
Finally, the middle group is directed to the inlier quantization path, while outliers, whether from inner or outer groups, are routed to the outlier quantization path, with zeros inserted in the alternate path accordingly.
%

%
Both the {\large\ding{173}} \textbf{inlier} and {\large\ding{174}} \textbf{outlier quantizer modules} handle quantization for each key and value vector.
Quantization scaling factors are dynamically computed during runtime, based on per-token min and max values for each group.
After calculating the scaling factors, these two modules finally perform 4/5-bit uniform quantization.
The quantized inlier and outlier values are merged using an \textsf{OR} gate and sent to the quantized dense matrix.
The outlier module generates an index and a group flag for COO transformation.
It employs a zero-remove shifter~\cite{rhu2018compressing, wang2021spatten} to implement fused dense-and-sparse encoding, optimizing memory usage.
\niparagraph{Dequantization engine.}
The dequantization engine, illustrated in Figure~\ref{fig-compression-engine}(b), is also integrated into the DMA unit to dequantize the KV cache retrieved from memory.
%

%
The {\large\ding{175}} \textbf{inlier dequantizer module} buffers incoming dense data to synchronize with sparse data processed by the outlier dequantizer module.
%
The {\large\ding{176}} \textbf{outlier dequantizer module} handles sparse COO data by performing a zero-insert~\cite{rhu2018compressing, wang2021spatten} operation to restore the original data alignment.
It identifies the original positions of fused outliers using the index and group information of the sparse data and inserts the necessary zeros accordingly.
Both dequantizer modules then restore the data, which is buffered to be aligned with the outputs of the counterpart module.
Finally, the outputs from both dequantizer modules are merged via an \textsf{OR} gate and forwarded to the processing unit.
Since the dequantization engine does not require the entire KV cache for its operation, we designed it to function in a streaming manner.
This design allows the dequantization engine to maintain low latency, while efficiently processing all past KV cache.

\begin{figure}[t]
    \centering
    \includegraphics[width=1.0\linewidth]{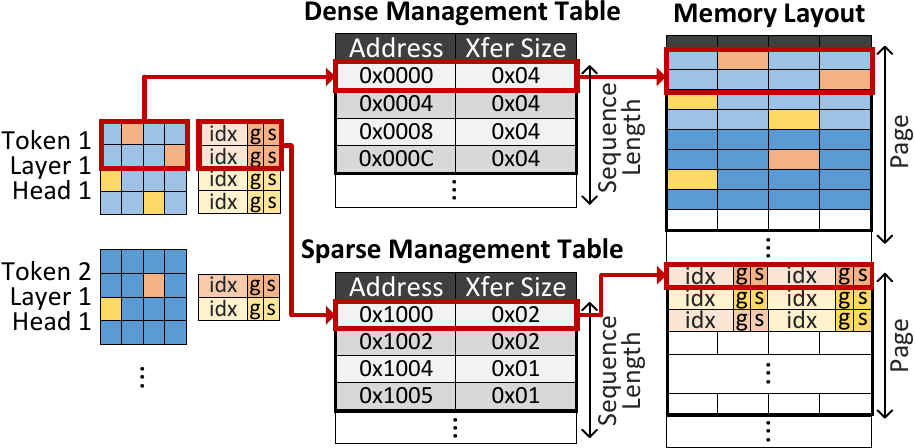}
    \vspace{-3.0ex}
    \caption{Operations of memory management unit (MMU) for handling dense and sparse data.}
    \Description{Operations of memory management unit (MMU) for handling dense and sparse data.}
    \label{fig-mem-layout} 
    \vspace{-2.0ex}
\end{figure}

\niparagraph{Memory management unit.} 
Figure~\ref{fig-mem-layout} illustrates the operations of memory management unit, which manages the reading and writing of quantized KV cache.
We design the MMU unit to handle dense-and-sparse matrices in a page-based manner, optimizing bandwidth utilization.
It supports multiple memory accesses in burst mode and streamlined operations to hide latency of memory and quantization/dequantization operations.
Since Oaken's MMU units share a common address space, MMU operates in each compute core independently, preventing interference.
Without this specialized MMU, processing variable-sized sparse matrices would require additional overhead for indexing, reshaping, and subsequent operations.
There are two major challenges associated with the design of MMU unit:

\begin{description}[labelindent=0.0em,nolistsep,leftmargin=1.0em]
\item[(1)]
\niparagraph{Addresses and transfer size management:}
Dense matrices have predictable sizes that are well-aligned within memory spaces, while sparse matrices vary in size.
Thus, management tables are needed for both dense and sparse data to accommodate this variability.
These tables contain the virtual-to-physical address mappings and transfer sizes for the KV cache, considering up to the maximum sequence length per attention head.
Physical addresses and transfer sizes are dynamically calculated during inference by checking available pages on demand.
\item[(2)]
\niparagraph{Read-write granularity and order determination:}
To maximize memory bandwidth utilization, burst access should be leveraged whenever possible to reduce the total number of memory transactions.
Writing KV cache involves relatively small sizes, as it only includes the key-value for the current token, whereas reading requires retrieving the KV cache for all previous tokens.
To address this, Oaken organizes KV cache for the current token in a layout that facilitates burst reads in subsequent operations.
Key-value vectors generated in the current layer are divided by attention head and written to distinct pages, as explained in Section~\ref{sec:compute-core}.
When the KV cache for the next token is generated, it is divided similarly and written sequentially, immediately following the previous tokens' KV cache.
This sequential arrangement allows that the KV cache for all previous tokens can be read in burst mode, allowing Oaken to efficiently utilize bandwidth.
\end{description}

\subsection{Token-level Batch Scheduling}
\label{sec:token-level-scheduling}
Efficient scheduling is crucial for Oaken to efficiently serve LLM inference.
Each compute core in Oaken is optimized to process a single token efficiently.
During the prefill phase, input tokens from each request are scheduled for parallel processing across multiple cores.
However, in the generation phase, each core handles a single output token from one request, which reduce hardware utilization.
For larger batches, Oaken improves overall core utilization by processing multiple requests in parallel.
Although the overhead for KV cache quantization and dequantization is minimal, Oaken further minimizes this by overlapping them with other operations.
In batched inference, KV cache cannot be shared across cores because each core processes distinct requests, forcing each core to monopolize the memory bandwidth.
Oaken employs a scheduling strategy that hides latency by overlapping KV quantization and dequantization with DMA reads and attention computations from other requests.

\section{Evaluation}
\label{sec:evaluation}

\begin{figure*}[t]
        \centering
        \includegraphics[width=1.0\linewidth]{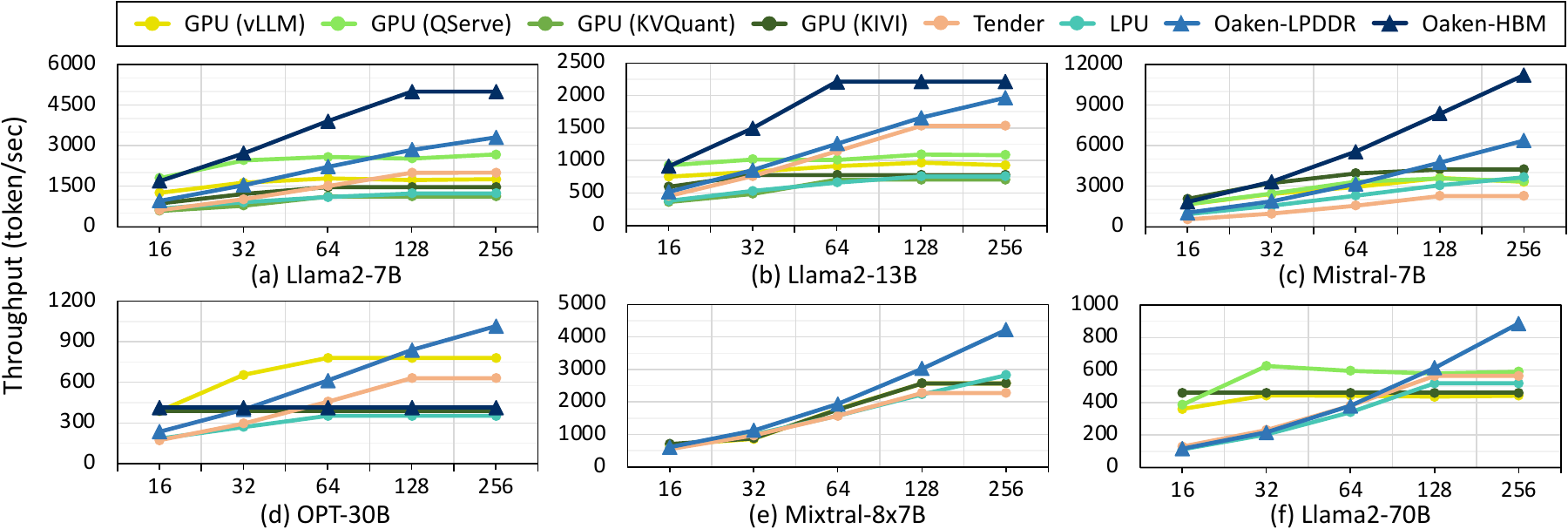}
        \vspace{-3ex}
        \caption{Throughput results of GPU baselines, LPU, Tender, and Oaken equipped with LPDDR and HBM across six LLMs. We sweep the batch size from 16 to 256. The input and output sequence lengths are set to 1K:1K.}
        \Description{Throughput results of GPU baselines, LPU, Tender, and Oaken equipped with LPDDR and HBM across six LLMs. We sweep the batch size from 16 to 256. The input and output sequence lengths are set to 1K:1K.}
        \label{fig:throughput}
        \vspace{+1ex}
\end{figure*}
\begin{table}[t]
    \caption{Hardware specification of NVIDIA A100 GPU and Oaken equipped with either HBM or LPDDR memory.}
    \Description{Hardware specification of NVIDIA A100 GPU and Oaken equipped with either HBM or LPDDR memory.}
    \vspace{-1.0ex}
    \centering
    \includegraphics[width=1.0\linewidth]{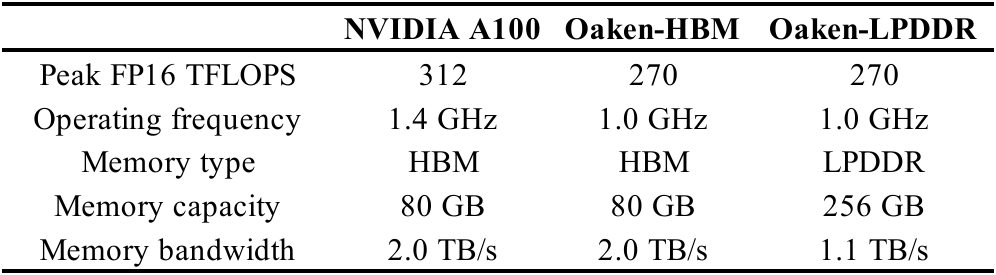}
    \label{tab-hw-metho} 
    \vspace{-3.0ex}
\end{table}

\subsection{Methodology}
\label{sec:methodology}
\niparagraph{Models and datasets.}
For evaluation, we use Llama2-7B, 13B, and 70B~\cite{touvron2023llama2}, OPT-6.7B, 13B, and 30B~\cite{zhang2022opt}, Mistral-7B~\cite{jiang2023mistral} and Mixtral-8x7B~\cite{jiang2024mixtralexperts} models.
Llama2, Mistral, and Mixtral models implement grouped-query attention~\cite{ainslie2023gqa}, while Mistral and Mixtral models also incorporate a sliding window~\cite{beltagy2020longformer-slidingwindow}.
Additionally, Mixtral model further integrates mixture-of-experts (MoE) layers~\cite{shazeer2017moe}.
To evaluate model accuracy, we utilize Wikitext2, PIQA, Winogrande, and Hellaswag datasets, which are widely evaluated in prior studies~\cite{lee2024owq, kim2023finequant, hooper2024kvquant, zhao2024alisa, lin2024qserve, zhao2024atom, lee2024tender}.
Wikitext2~\cite{merity2016pointer} dataset consists of tokens extracted from Wikipedia articles, while PIQA~\cite{bisk2019piqa}, Winogrande~\cite{sakaguchi2019winogrande}, and Hellaswag~\cite{zellers2019hellaswag} constitute questions and answers.
We report zero-shot accuracy (\%) for PIQA, Winogrande, and Hellaswag datasets and perplexity for Wikitext2 dataset.
Note that for perplexity, lower values indicate better performance.
For real-world benchmarking, we use two open-source production traces from Azure LLM inference services, \textit{Conversation}~\cite{patel2024splitwise, azure-inference-dataset} and \textit{BurstGPT}~\cite{burstgpt}.
We follow the methodology established in prior work to simulate inference serving scenarios~\cite{heo2024neupims}.
Requests are sampled from the trace over a time period, and batches are synthesized with varying input and output sequence lengths.
We repeat this process across multiple batches, measuring the average performance.

\begin{table*}[t]
        \centering
        \caption{Perplexity results on Wikitext2, and zero-shot accuracy results on PIQA, Winogrande, and Hellaswag datasets with effective bitwidth of each quantization technique.}
        \Description{Perplexity results on Wikitext2, and zero-shot accuracy results on PIQA, Winogrande, and Hellaswag datasets with effective bitwidth of each quantization technique.}
        \vspace{-1.0ex}
        \includegraphics[width=1.0\linewidth]{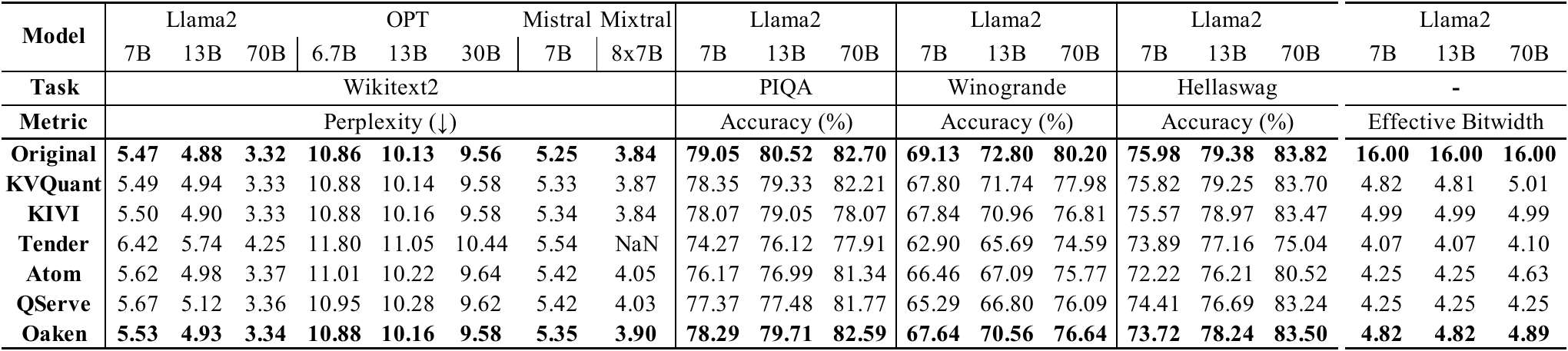}
        \label{tab:accuracy}
        \vspace{-1.0ex}
\end{table*}

\niparagraph{Accelerator platforms.}
For the end-to-end performance evaluation, we developed a hardware simulator for the Oaken accelerator by extending the existing hardware simulator of LPU~\cite{hong2022dfx, park2024lpddr}.
LPU was initially optimized for low-latency inference without batching support, while the follow-up work scaled it to accommodate larger batches~\cite{moon2024lpu}.
We extend the LPU architecture to integrate Oaken by incorporating quantization/dequantization engines and memory management units into the LPU's DMA units.
For GPU baselines, we use NVIDIA A100 GPUs equipped with 80 GB HBM~\cite{nvidia-a100}.
We use a single GPU for Llama2-7B, 13B, and Mistral-7B models, as it could accommodate all the model parameters.
For larger models including OPT-30B, Mixtral-8x7B, and Llama2-70B models, we use two GPUs, employing pipeline parallelism to keep computation capability and memory bandwidth consistent, while scaling capacity to 160 GB.

\niparagraph{Hardware specifications and implementation.}
Table~\ref{tab-hw-metho} summarizes the specifications of the NVIDIA A100 GPU and the Oaken accelerator used in our evaluations.
Oaken-LPDDR is equipped with an LPDDR memory module matching the specifications used in prior works~\cite{park2024lpddr, LLMCompass}.
Oaken-HBM is configured with HBM memory identical to that of the A100 GPU.
The HBM memory offers higher bandwidth but has a smaller capacity compared to LPDDR memory, making a trade-off between them.
We implement the Oaken hardware in RTL using SystemVerilog and verify its functionality with Synopsys VCS functional verification solution.
The RTL is synthesized using Synopsys Design Compiler for a target clock frequency of 1 GHz on TSMC 28nm technology.

\niparagraph{Baselines.}
For accuracy evaluation, we use Tender~\cite{lee2024tender}, Atom~\cite{zhao2024atom}, QServe~\cite{lin2024qserve}, KIVI~\cite{liu2024kivi}, and KVQuant~\cite{hooper2024kvquant} as baselines.
For performance evaluation, we first use vLLM~\cite{vllm-2023} as the FP16-operating GPU baseline, as it represents the state-of-the-art LLM serving system and delivers superior performance compared to other alternatives.
We also use most of the baselines for accuracy evaluation, excluding Atom, which lacks open-source code availability.
KIVI, QServe, and KVQuant serve as additional GPU baselines, running on A100 GPUs.
Tender~\cite{lee2024tender} is an LLM inference accelerator employing quantization, which offers an open-source simulator.
For a fair comparison, we align Tender's memory specifications and compute capabilities with those of the A100 GPU.
All quantization-based baselines employ 4-bit KV cache-only quantization.
While QServe and Tender offer weight and activation quantization, we disable these features for fair comparison with the other baselines.

\niparagraph{Thresholds.}
Throughout the evaluation, we set the outer, middle, and inner group ratio to 4\%, 90\%, and 6\%, respectively.
This global configuration applies to all models and datasets for the following two reasons.
First, as discussed in Section~\ref{sec:kv-distribution}, KV cache distribution is independent of the input dataset.
Second, although the optimal group ratio varies slightly across LLMs, its impact on inference performance and accuracy is marginal.
Section~\ref{sec:eval-result} explores the threshold search space and group count, justifying this choice.

\niparagraph{Offline profiling.}
Oaken's offline profiling is performed by collecting \textsf{topK} values, which represent four boundaries of the quantization groups, during inference and averaging the gathered values.
As mentioned, we use the same group ratio and the Wikitext2 dataset for all LLMs.
However, since the KV cache distribution varies by model, requiring different group thresholds and individual profiling.
Despite this, Oaken's offline profiling requires only about a hundred inferences and takes approximately ten minutes, even for the Llama2-70B model.
Since this process is required only once before serving LLM inference online, the overhead is negligible.

\subsection{Experimental Results}
\label{sec:eval-result}
\niparagraph{Throughput.}
Figure~\ref{fig:throughput} presents the end-to-end throughput comparison results among GPU baselines (vLLM, KVQuant, KIVI, and QServe) and ASIC accelerators (Tender and Oaken).
The results are omitted when the baseline system lacks support for the corresponding models. 
For a batch size of 256, Oaken-LPDDR achieves an average throughput improvement of 1.79$\times$ over vLLM and 1.58$\times$ over QServe.
This improvement is attributed to the reduced execution time of the attention operations, which accounts for the majority of inference time.
Oaken alleviates bandwidth and capacity bottlenecks by minimizing memory access to the KV cache.

GPU baselines perform well for small batches and models, but as the batch size grows, they cannot accommodate the entire batch due to capacity constraints, leading to performance saturation.
Tender, which uses the same HBM memory as A100 GPUs, also does not scale for large batches.
Oaken-HBM outperforms other baselines and Oaken-LPDDR for small models and batch sizes.
However, it faces challenges in accommodating large models such as Mixtral-8x7B and Llama2-70B, or handling large batches due to its insufficient memory capacity.
Mistral-7B, Mixtral-8x7B, and Llama2-70B models employ grouped-query attention to reduce KV cache size, helping to alleviate bandwidth bottlenecks even without KV quantization.
However, for larger batch sizes, capacity limitations still cause saturation, while Oaken accelerators offer scalability, demonstrating the effectiveness of our KV cache quantization technique.

\niparagraph{Accuracy.}
Table~\ref{tab:accuracy} presents the accuracy results of each baselines across eight LLMs on Wikitext2, PIQA, Winogrande, and Hellaswag datasets, along with the effective bitwidth on the Llama2 models.
Oaken exhibits an average accuracy loss of 0.87\% compared to the original FP16 baseline, with 0.54\% and 0.32\% lower accuracy than KVQuant and KIVI, respectively, while achieving 1.38\% higher accuracy than QServe.
KVQuant and KIVI requires a larger effective bitwidth due to their use of sparse layout for outlier values and fine-grained grouping, respectively.
They achieve higher accuracy than Oaken, but their advantages are largely offset by the prohibitively high overhead of online sorting and mixed-precision operations.
On the other hand, Tender, Atom, and QServe employ an indirect indexing technique and a transformation matrix to reorder KV channels and group those with similar magnitudes.
This approach impose minimal overhead due to their low effective bitwidth, as they do not require individual processing of outliers; however, this comes at the cost of larger accuracy losses, as they rely on coarse-grained per-group or per-channel quantization without considering exceptions in KV distribution, as discussed in Section~\ref{sec:kv-distribution}.
%

\begin{figure}[t]
    \centering
    \includegraphics[width=1.0\linewidth]{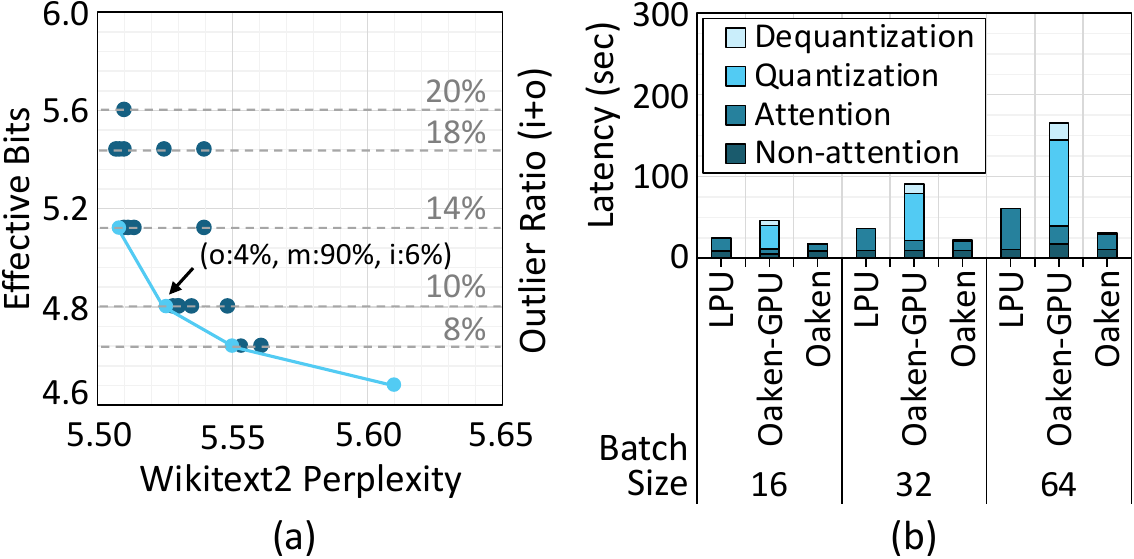}
    \vspace{-3ex}
    \caption{(a) Accuracy and effective bits with varying quantization group ratios. (b) Latency breakdown for non-attention, attention, quantization and dequantization operations using Llama2-7B model across varying batch size.}
    \Description{(a) Accuracy and effective bits with varying quantization group ratios. (b) Latency breakdown for non-attention, attention, quantization and dequantization operations using Llama2-7B model across varying batch size.}
    \label{fig:latency-breakdown}
    \vspace{-2.0ex}
\end{figure}

\niparagraph{Trade-off between accuracy and compression ratio.}
Figure~\ref{fig:latency-breakdown}(a) illustrates the trade-off between accuracy and compression ratio in Oaken's KV cache quantization on Llama2-7B model, which can also be interpreted as a trade-off between accuracy and performance.
We sweep the group ratio and measure the perplexity on Wikitext2 dataset.
The effective bits in Oaken are determined by the ratio of inner and outer groups, as their location is stored in the sparse matrix.
All points on the same horizontal line share the same outlier ratio and effective bits, but differ in group composition.
We select a ratio of 4\%, 90\%, and 6\% for the outer, middle, and inner groups, respectively, throughout the entire evaluation, as they are one of the Pareto-optimal points highlighted as a light blue line.
While using higher effective bits might improve accuracy even better than the current configuration, it negatively impacts the inference performance due to its low compression ratio.
Table~\ref{tab:group_sweep} presents the perplexity and effective bitwidth for different numbers of groups evaluated on the Llama2-7B model.
We fix the total ratio of inner and outer groups at 10\%.
Oaken's fused dense-and-sparse encoding eliminates the need for a bit to represent outliers when using two groups.
However, this disrupts memory alignment, as each sparse COO entry consists of 6 index bits and 1 sign bit.
To mitigate hardware overhead, an extra padding bit is added, maintaining the same effective bitwidth.
Using four or five groups improves accuracy but increases bitwidth, as 9-bit COO entries require two bits for inner and outer groups.
This also misaligns memory layout, requiring additional padding.
While using 4-bit outliers to keep 8-bit alignment preserves the effective bitwidth, it slightly reduces accuracy.
In summary, Oaken's three-group quantization offers the optimal balance between cost and accuracy.

\niparagraph{Latency breakdown.}
To better understand the impact of KV cache quantization on performance, we break down the end-to-end inference latency of LPU and Oaken-LPDDR as we vary the batch size.
We also implement Oaken's quantization algorithm on GPU and measure its operation latencies.
Figure~\ref{fig:latency-breakdown}(b) shows that the latency of attention operations increases proportionally with the batch size.
While Oaken does not directly affect the execution time of non-attention operations, it alleviates bandwidth bottleneck through KV quantization.
As a result, the execution time of attention operations is, on average, 55.0\% shorter than that of LPU, contributing to a reduction in end-to-end latency.
When the batch size is 64, quantization and dequantization account for only 1.29\% and 3.23\% of the entire latency, respectively.
On the contrary, Oaken algorithm on GPU demonstrates long quantization and dequantization latencies due to warp divergence in CUDA, which is required to separate multiple quantization groups.
Note that Oaken hides both quantization and dequantization latencies by overlapping them with other operations and processing them in a streaming manner.

\begin{table}[t]
    \centering
    \caption{Accuracy and effective bits using the Llama2-7B model with varying number of groups and group ratios while keeping the total inner and outer group ratio at 10\%.}
    \Description{Accuracy and effective bits using the Llama2-7B model with varying number of groups and group ratios while keeping the total inner and outer group ratio at 10\%.}
    \vspace{-1.0ex}
    \includegraphics[width=0.80\linewidth]{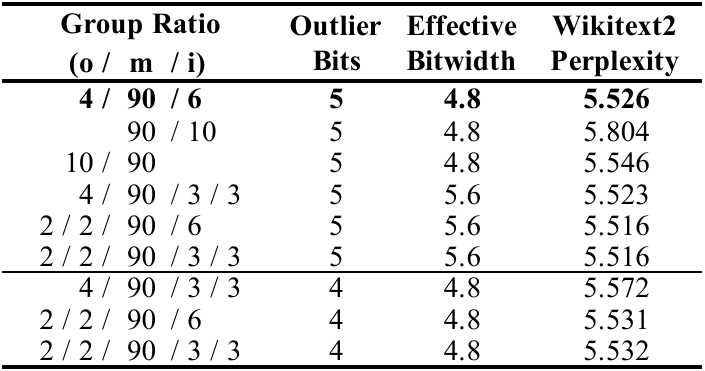}
    \label{tab:group_sweep}
    \vspace{-0.0ex}
\end{table}

\begin{figure}[t]
        \centering
        \includegraphics[width=1.0\linewidth]{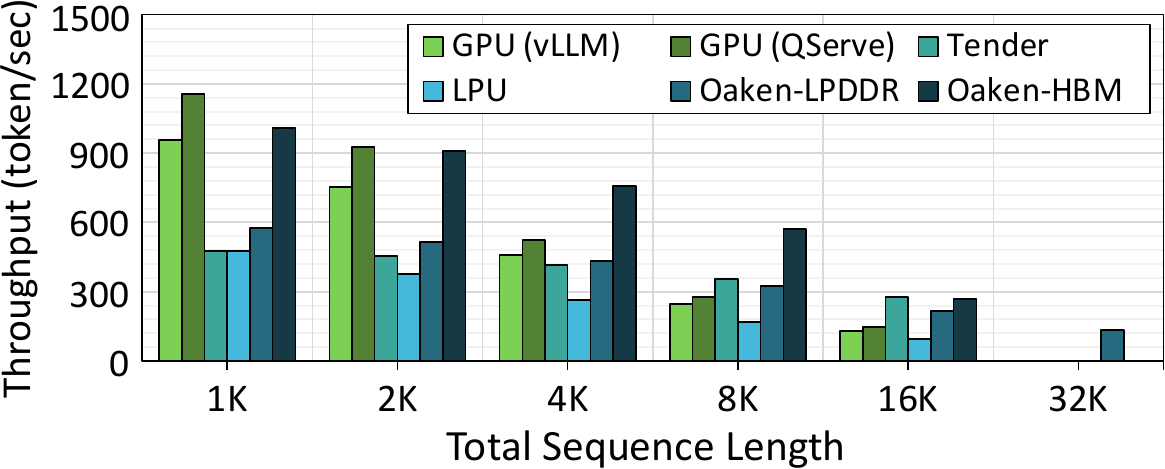}
        \vspace{-4.0ex}
        \caption{Throughput results on Llama2-13B model with a batch size of 16 when increasing total sequence length from 1K to 32K. The ratio of input and output length is set to 1:1.}
        \Description{Throughput results on Llama2-13B model with a batch size of 16 when increasing total sequence length from 1K to 32K. The ratio of input and output length is set to 1:1.}
        \label{fig:seq-len-sensitivity}
        \vspace{-0.0ex}
\end{figure}

\niparagraph{Sensitivity to sequence length.}
Figure~\ref{fig:seq-len-sensitivity} shows throughput results when sweeping the total sequence length from 1K to 32K.
For shorter sequence lengths below 8K, the proportion of compute-bound, batchable operations is larger than memory-bound, \emph{non}-batchable attention operations.
Therefore, the performance of QServe and vLLM outperform Oaken in this range by leveraging the higher parallelizable resources available on GPUs.
However, as the sequence length increases, Oaken-HBM surpasses other baselines, including Oaken-LPDDR, with its high memory bandwidth and KV cache quantization.
However, HBM-based systems including QServe and Oaken-HBM cannot handle sequences longer than 16K, making it difficult to complete the entire batch due to insufficient capacity.
Oaken-LPDDR, on the other hand, can accommodate longer sequences of up to 32K by mitigating both bandwidth and capacity pressure through KV quantization and large-capacity memory.

\begin{figure}[t]
        \centering
        \includegraphics[width=1.0\linewidth]{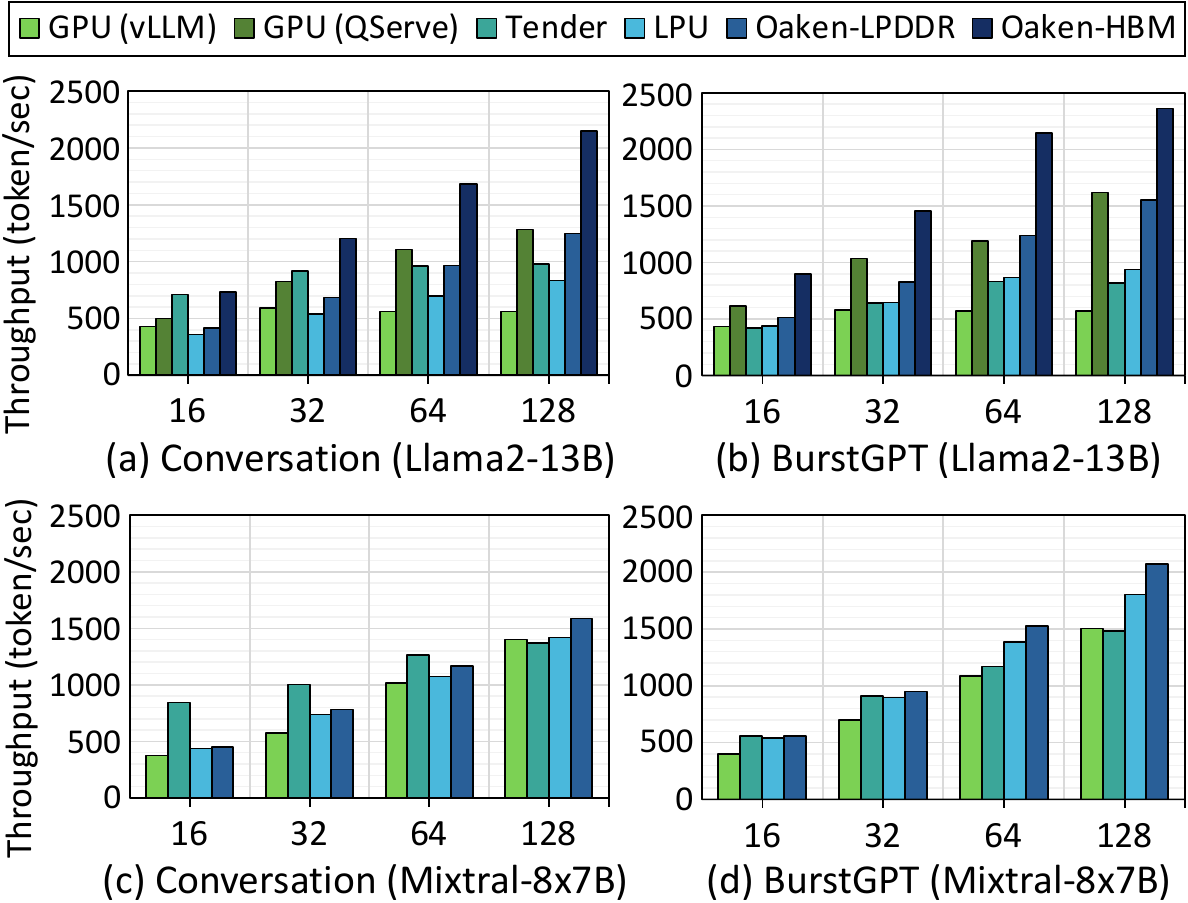}
        \vspace{-2.0ex}
        \caption{Real-world benchmark results for generation throughput on Llama2-13B and Mixtral-8x7B models evaluated with two LLM inference traces: \textit{Conversation}~\cite{azure-inference-dataset} and \textit{BurstGPT}~\cite{burstgpt}. We sweep the batch size from 16 to 128.}
        \Description{Real-world benchmark results for generation throughput on Llama2-13B and Mixtral-8x7B models evaluated with two LLM inference traces: Conversation and BurstGPT. We sweep the batch size from 16 to 128.}
        \label{fig:application-benchmark}
        \vspace{-0.0ex}
\end{figure}

\niparagraph{Real-world benchmark.}
Figure~\ref{fig:application-benchmark} presents the generation throughput results for Llama2-13B and Mixtral-8x7B models, evaluated using two real-world LLM inference traces.
We exclude Oaken-HBM and QServe for Mixtral-8x7B model, as Oaken-HBM's memory cannot accommodate the entire model and QServe lacks support for MoE layers.
Tender, which employs systolic arrays, suffers underutilization due to the padding required by varying prompt lengths within a batch.
\textit{Conversation} trace features short output lengths, resulting in a brief generation phase.
As the bandwidth bottleneck due to the KV cache is noticeable in generation phase, a short generation length reduces the advantage of Oaken's KV cache quantization.
%
%
Conversely, \textit{BurstGPT} trace features longer output lengths, where KV cache quantization in Oaken becomes more beneficial.
Mixtral-8x7B model utilizes grouped query attention to reduce its KV cache size compared to multi-head attention. 
Quantization baselines, including Oaken-LPDDR and Tender, show little to no performance gain over full-precision baselines.
However, as batch size increases or with the BurstGPT trace with longer generation lengths, Oaken-LPDDR demonstrates greater performance gains.
In summary, Oaken delivers an advantage over existing solutions in real-world scenarios for the Llama2-13B and Mixtral-8x7B models.

\niparagraph{Area and power.}
As reported in Table~\ref{tab-area}, the quantization and dequantization engines account for a minor percentage of the \emph{total compute core} area at 1.86\% and 6.35\%, respectively.
In addition, the power consumption of the entire accelerator embedded with Oaken modules is 222.7W, which is 44.3\% lower than the 400W TDP of the A100 GPU.
These results clearly indicate that integrating Oaken's quantization and dequantization modules only imposes minimal hardware overhead while improving performance and achieving better energy efficiency compared to the GPU. 

\section{Related Work}

\niparagraph{LLM quantization.}
Most prior work on LLM quantization focuses on weight and/or activation quantization~\cite{lin2023awq, shao2023omniquant, frantar2022optq, wei2022outlier, wei2023outlier, liu2023qllm} to reduce inference computation costs.
RPTQ~\cite{yuan2023rptq}, SpinQuant~\cite{liu2024spinquant}, and QuaRot~\cite{ashkboos2024quarot} introduce transformation matrices for weight and activation quantization.
SmoothQuant~\cite{xiao2023smoothquant} mitigates the quantization difficulty by transferring activation outlier scales to weights.
SqueezeLLM~\cite{kim2023squeezellm} applies dense-and-sparse quantization for storing weight outliers in full precision.
However, many existing approaches overlook the KV cache, whose size scales with sequence length and batch size, often becoming a major bottleneck for latency and throughput in batched LLM inference.
Oaken overcomes this issue by employing an offline-online KV cache quantization algorithm with a customized hardware module, achieving high throughput with minimal accuracy degradation.
%

%

\begin{table}[t] 
    \caption{Area overhead analysis of compression and decompression engines on TSMC 28nm.}
    \Description{Area overhead analysis of compression and decompression engines on TSMC 28nm.}
    \vspace{-1.0ex}
    \centering
    \includegraphics[width=0.95\linewidth]{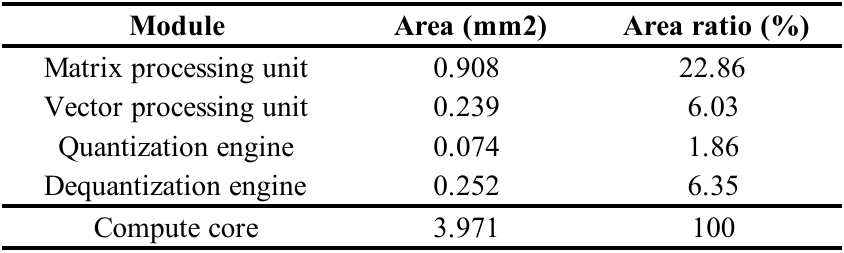}
    \label{tab-area} 
    \vspace{-0.0ex}
\end{table}

%
\niparagraph{LLM inference accelerator.}
DFX~\cite{hong2022dfx} is one of the first LLM accelerators, which is designed to accelerate the entire GPT2 model using HBM and FPGA.
CXL-PNM~\cite{park2024lpddr} introduces a LLM accelerator largely leveraging the DFX design, while employing LPDDR for striking a sweet spot in the bandwidth-capacity tradeoff space for large-scale LLM serving. 
LPU~\cite{moon2024lpu} is another LLM accelerator that differs from prior work in optimizing its design for minimal latency.
TransPIM~\cite{transpim22} proposes a PIM accelerator targeted for encoder-based transformer models such as BERT.
AttAcc~\cite{attacc-asplos24}, IANUS~\cite{ianus-asplos24}, and NeuPIMS~\cite{heo2024neupims} employ PIM technologies for decoder-based transformer LLM serving, while they impose capacity pressure on the large-batch long-sequence serving scenarios. 
Unlike these prior works, this work devises Oaken, which jointly employs KV quantization and LPDDR for unleashing larger capacity and increased bandwidth to enable fast and efficient LLM serving.
\niparagraph{Acceleration for quantized model inference.}
There have been several prior works on accelerating quantized neural network inference~\cite{qin2023fact, zeng2024flightllm, mixgemm-hpca23, mix-and-match, drq-isca20,inci2023quidam, keller2023vsq, yang2022gqna, Li2024mvm, shen2024agile}.
Mokey~\cite{zadeh2022mokey} and Olive~\cite{guo2023olive} apply outlier-aware quantization methods to transformer-based LLMs.
LUT-GEMM~\cite{park2023lutgemm} proposes a lookup-table-based GPU kernel to eliminate dequantization overhead in quantized LLM inference.
Tender~\cite{lee2024tender} quantizes KV cache as well as weights and activations, but its accuracy loss is significant.
In contrast, Oaken is an LLM inference acceleration solution optimized for KV cache quantization, offering a scalable solution while minimizing accuracy loss. 
\section{Conclusion}

Batched LLM inference faces significant challenges from high memory bandwidth and capacity demands, exacerbated by the growing size of KV caches in modern LLMs that produce long-sequence outputs. 
This paper tackles this challenge by proposing an acceleration solution, Oaken, that jointly exploits (1) an offline-online hybrid KV cache compression technique and (2) custom hardware modules tailored for the proposed algorithm that can be integrated with LLM accelerators.
Oaken effectively unlocks sufficient bandwidth and capacity, which would otherwise be unattainable, leading to significant throughput improvements with only marginal accuracy loss.
These compelling advantages demonstrate that Oaken efficiently addresses the two primary bottlenecks of modern LLM serving.  
%



\begin{acks}
%






%
We thank the anonymous reviewers for their comments and feedback.
This work was supported by the Institute of Information \& Communications Technology Planning \& Evaluation (IITP) (No.2018-0-00503, No.RS-2024-00459797, Development of ML compiler framework for on-device AI), IITP under the Graduate School of Artificial Intelligence Semiconductor (IITP-2025-RS-2023-00256472), and the National Research Foundation of Korea (NRF) (RS-2024-00342148), grant funded by the Korea government (MSIT).
This work was also partly supported by HyperAccel.
\end{acks}

\balance
\bibliographystyle{ACM-Reference-Format}
\bibliography{refs}

\end{document}